\documentclass[12pt]{article}

\usepackage{array} 
\usepackage{amssymb}
\usepackage{graphics,graphpap}
\usepackage{graphicx}
\usepackage{color}
\usepackage{graphicx}
\usepackage{dcolumn}
\usepackage{epsfig}
\usepackage{epstopdf}
\usepackage{bbm}
\usepackage{amsmath}
\usepackage{amsfonts}
\usepackage{textcomp}
\usepackage{setspace}

\newcommand{\beq}{\begin{eqnarray}}
\newcommand{\eeq}{\end{eqnarray}}

\newcommand{\bmp}{\noindent\begin{minipage}{16cm}}
\newcommand{\emp}{\end{minipage}\vskip 7mm} 

\def\drawbox#1#2{\hrule height#2pt
        \hbox{\vrule width#2pt height#1pt \kern#1pt
              \vrule width#2pt}
              \hrule height#2pt}

\def\Asym#1#2{\vcenter{\vbox{\drawbox{#1}{#2}
              \kern-#2pt 
              \drawbox{#1}{#2}}}}

\def\simge{\mathrel{%
   \rlap{\raise 0.511ex \hbox{$>$}}{\lower 0.511ex \hbox{$\sim$}}}}

\def\simle{\mathrel{
   \rlap{\raise 0.511ex \hbox{$<$}}{\lower 0.511ex \hbox{$\sim$}}}}

\def\s#1{\setbox0=\hbox{$#1$}%
\rlap{\ifdim\wd0>.7em\kern.22\wd0\else\kern.1\wd0\fi /}#1}

\begin{document}

\begin{titlepage}
\title{\vspace*{-2.0cm}
\bf\Large
Constraining models for keV sterile neutrinos by quasi-degenerate active neutrinos\\[5mm]\ }

\author{
Alexander Merle\thanks{email: \tt A.Merle@soton.ac.uk}
\\ \\
{\normalsize \it Physics and Astronomy, University of Southampton,}\\
{\normalsize \it Southampton, SO17 1BJ, United Kingdom}\\
}
\date{\today}
\maketitle
\thispagestyle{empty}

\begin{abstract}
\noindent
We present a No-Go theorem for keV sterile neutrino Dark Matter: if sterile neutrinos at the keV scale play the role of Dark Matter, they are typically unstable and their decay produces an astrophysical monoenergetic X-ray line. It turns out that the observational bound on this line is so strong that it contradicts the existence of a quasi-degenerate spectrum of active neutrinos in a seesaw type~I framework where the Casas-Ibarra matrix $R$ is real. This is the case in particular for models without $CP$ violation. We give a general proof of this theorem. While the theorem (like every No-Go theorem) relies on certain assumptions, the situation under which it applies is still sufficiently general to lead to interesting consequences for keV neutrino model building. In fact, depending on the outcome of the next generation experiments, one might be able to rule out whole classes of models for keV sterile neutrinos.
\end{abstract}

\end{titlepage}

\section{\label{sec:intro}Introduction}

In spite of having considerable astrophysical evidence for Dark Matter (DM)~\cite{Komatsu:2010fb}, we still do not know its true identity. While the condition to correctly form cosmological structure rules out hot (highly relativistic) DM~\cite{Abazajian:2004zh,dePutter:2012sh}, the other two possibilities of cold (non-relativistic) or warm (moderately relativistic) DM are still consistent with all constraints. Apart from weakly interacting massive particles (WIMPs), which are generic cold DM candidates, the case of particle warm or cold DM with a mass of a few keV has attracted the interest of the structure formation simulation community~\cite{Bode:2000gq,Hansen:2001zv,Boyarsky:2008xj,Lovell:2011rd,Boyanovsky:2010pw,Boyanovsky:2010sv,VillaescusaNavarro:2010qy}, too. In addition, this was fueled by model-independent analyses~\cite{deVega:2009ku,deVega:2010yk} and surveys like ALFALFA~\cite{Papastergis:2011xe}, which suggest the possibility of keV-DM.

One of the prime candidates for keV-scale DM with suppressed interactions is a sterile neutrino $N_1$. The main framework proposed is the $\nu$MSM~\cite{Asaka:2005an}, which can simultaneously account for neutrino oscillations, Dark Matter, and the baryon asymmetry of the Universe~\cite{Canetti:2012vf,Canetti:2012kh}. In fact, sterile (or right-handed) neutrinos are necessary in many frameworks with a non-zero light neutrino mass, which makes it attractive to use them not only for the generation of a small active neutrino mass but also for other purposes. Furthermore, such settings offer the possibility to probe the DM sector with the help of low-energy neutrino data, provided that the two are tightly linked in a concrete model.

Accordingly, keV sterile neutrinos have been investigated in many contexts, such as Left-Right symmetry~\cite{Bezrukov:2009th,Nemevsek:2012cd}, composite neutrinos~\cite{Grossman:2010iq,Robinson:2012wu}, or within the scotogenic model~\cite{Sierra:2008wj,Gelmini:2009xd}. The two main tasks are to motivate the existence of the keV scale in the first place, and to generate the correct DM relic abundance, see Ref.~\cite{Abazajian:2012ys} for an overview. Several models that achieve the former point have been suggested, using a flavour symmetry~\cite{Shaposhnikov:2006nn,Lindner:2010wr,Araki:2011zg,Allison:2012qn}, the Froggatt-Nielsen mechanism~\cite{Barry:2011wb,Merle:2011yv,Barry:2011fp}, split seesaw~\cite{Kusenko:2010ik,Adulpravitchai:2011rq}, or an extended seesaw mechanism~\cite{Barry:2011wb,Zhang:2011vh}. Examples for production mechanisms are resonant~\cite{Shi:1998km} and non-resonant~\cite{Dodelson:1993je} (non-) thermal production, in particular the decay of a scalar particle~\cite{Shaposhnikov:1900zz}, or thermal overproduction with subsequent entropy dilution~\cite{Scherrer:1984fd}. These ideas have been applied to keV sterile neutrinos or to extended scanerios (see, e.g., Refs.~\cite{Bezrukov:2009th,Nemevsek:2012cd,Asaka:2006ek,Asaka:2006rw,Asaka:2006nq,Laine:2008pg,Wu:2009yr,Shaposhnikov:2006xi,Bezrukov:2012as,Khalil:2008kp,King:2012wg}).

Models of the flavour structure are very attractive to explain the patterns in the light (active) neutrino spectrum. While by neutrino oscillation experiments we have measured all leptonic mixing angles and mass square differences~\cite{Tortola:2012te,Fogli:2012ua,GonzalezGarcia:2012sz}, we still do not know the absolute neutrino mass scale.  On the other hand, from kinematical determinations of the neutrinos mass~\cite{Lobashev:2000vb,Kraus:2004zw}, from experiments on neutrinoless double beta decay~\cite{KlapdorKleingrothaus:2000sn,Andreotti:2010vj}, and from cosmological observations~\cite{Komatsu:2010fb}, it is fair to say that the true neutrino mass is below $1$~eV. However, we still do not know the pattern of the light neutrino masses, they could be either normally ($m_1 < m_2 < m_3$) or inversely ($m_3 < m_1 < m_2$) ordered. In case the absolute light neutrino mass scale $m_0$ is much larger than the corrections by the two mass-square differences, $m_0 \gg \sqrt{\Delta m_{A,\odot}^2}$, all light neutrino masses would be roughly equal, $m_{1,2,3} \simeq m_0$, and the corresponding pattern is called \emph{quasi-degenerate} (QD).

In this paper we present a little No-Go theorem, which proves that a QD mass pattern for light neutrinos is inconsistent with the existence of keV sterile neutrino DM in the absence of $CP$ violation or, even more general, in the case of a real Casas-Ibarra (CI) matrix $R$. After presenting and proving the theorem in Sec.~\ref{sec:theorem}, we give a discussion of its implications in Sec.~\ref{sec:disc} before concluding in Sec.~\ref{sec:conc}.

The work presented here is a generalization of the considerations in Refs.~\cite{Asaka:2005an,Boyarsky:2006jm}. Although in particular Ref.~\cite{Asaka:2005an} arrives at similar formulas, it for example does not contain a discussion on the aspects concerning CP violation. This is an important and fundamental point which becomes most easily visible in the CI parametrization: this parametrization separates the CP violation in the light neutrino sector, contained in the Pontocorvo-Maki-Nagakawa-Sakata (PMNS) matrix $U_{\rm PMNS}$, from the one in the heavy sector, contained in the CI matrix $R$. In particular when investigating a concrete model it is often possible to derive an explicit expression for $R$, which can immediately be used to match it with the considerations in this paper. An equally useful point presented here is the detailed discussion of when the No-Go theorem is \emph{not} applicable: when developing a new model, it may for certain reasons be desirable to avoid this constraint. This can be easily done by making use of the list of ways around the theorem presented here. Another new point is the connection to the cosmological sum $\Sigma$ of light neutrino masses. This is particularly important in the light of new cosmological data to be expected from the Planck satellite~\cite{Planck}, which is not unlikely to measure the value of $\Sigma$. Finally, the GERDA experiment~\cite{GERDA} can be expected to yield results soon~\cite{nuGERDA}, which makes the research presented here even more timely: in principle, GERDA could determine the light neutrino spectrum to be QD, by observing a corresponding rate of neutrino-less double beta decay, or it could strongly constrain the QD pattern for the case of Majorana neutrinos.

\section{\label{sec:theorem}The No-Go theorem}

In this section, we will show under which circumstances keV sterile neutrino Dark Matter is not consistent with the existence of quasi-degenerate active neutrinos. Let us state the theorem first and then give a proof of its validity.\\

\noindent {\bf No-Go theorem}:\\
Consider a type~I seesaw situation with three left-handed and three right-handed neutrinos, where the Casas-Ibarra matrix $R$ is real and where the lightest sterile neutrino mass eigenstate $N_1$ mainly decays via the channel $N_1 \to \nu \gamma$. Then, $N_1$ cannot play the role of keV Dark Matter while at the same time the light neutrino mass pattern is quasi-degenerate.\\

\noindent {\bf Proof}:\\
The usual $6\times 6$ mass matrix in a seesaw type~I situation is given by
\begin{equation}
 M_\nu = \begin{pmatrix}
 0 & m_D\\
 m_D^T & M_R
 \end{pmatrix}.
 \label{eq:p_1}
\end{equation}
In the charged lepton mass basis, the mass matrix $M_\nu$ is fully diagonalized by a unitary matrix $\overline{U}$, according to
\begin{eqnarray}
 && D \equiv \overline{U}^T M_\nu \overline{U} = \begin{pmatrix}
 D_\nu & 0\\
 0 & D_N
 \end{pmatrix},\ \ {\rm where} \label{eq:p_2} \\
 && \overline{U}\simeq \begin{pmatrix}
 \mathbbm{1} - \frac{1}{2} m_D^* {M_R^{-1}}^* M_R^{-1} m_D^T & m_D^* {M_R^{-1}}^*\\
 - M_R^{-1} m_D^T & \mathbbm{1} - \frac{1}{2} M_R^{-1} m_D^T m_D^* {M_R^{-1}}^*
 \end{pmatrix} \begin{pmatrix}
 U_{\rm PMNS} & 0\\
 0 & V_R
 \end{pmatrix},
 \nonumber
\end{eqnarray}
with $D_\nu = {\rm diag}(m_1, m_2, m_3) = U_{\rm PMNS}^T m_\nu U_{\rm PMNS}$ and $D_N = {\rm diag}(M_1, M_2, M_3) = V_R^T M_R V_R$. Here, $m_\nu = - m_D M_R^{-1} m_D^T$ is the usual seesaw type~I~\cite{Minkowski:1977sc,Yanagida:1979as,GellMann:1980vs,Glashow:1979nm,Mohapatra:1979ia} light neutrino mass matrix, with eigenvalues $m_{1,2,3}$. In turn, $M_R$ has eigenvalues $M_{1,2,3}$, where $0 < M_1 \ll M_{2,3}$.\footnote{One could ask whether the seesaw formula is at all applicable, since one might be in danger to divide by a small mass $M_1 = \mathcal{O}({\rm keV})$. However, anticipating the X-ray bound, cf.\ discussion before Eq.~\eqref{eq:p_10}, as well as the Lyman-$\alpha$ bound, cf.\ Sec.~\ref{sec:disc}, one can conclude that $|{m_D}_{e 1}|^2 + |{m_D}_{\mu 1}|^2 + |{m_D}_{\tau 1}|^2 \ll M_1^2$, and hence, due to the sum over absolute values, $|{m_D}_{\alpha 1}| \ll M_1$, which makes the seesaw formula valid. This behavior could be dubbed as the \emph{keV seesaw practicality theorem}.} The first matrix factor in $\overline{U}$ from Eq.~\eqref{eq:p_2} approximately block-diagonalizes the whole matrix $M_\nu$, while the unitary PMNS matrix $U_{\rm PMNS}$ and the unitary matrix $V_R$ diagonalize the submatrices $m_\nu$ and $M_R$, respectively.

We can hide our lack of knowledge on the $6\times 6$ matrix $M_\nu$ by shifting all unknowns into a complex orthogonal matrix $R$. This is called the \emph{Casas-Ibarra parametrization}~\cite{Casas:2001sr}, and in a basis where $M_R = {\rm diag}(M_1, M_2, M_3)$ it amounts to rewriting the Dirac mass matrix $m_D$ as
\begin{equation}
 m_D = i U^* {\rm diag}(\sqrt{m_1}, \sqrt{m_2}, \sqrt{m_3}) R^T {\rm diag}(\sqrt{M_1}, \sqrt{M_2}, \sqrt{M_3}),
 \label{eq:p_3}
\end{equation}
where $U\equiv U_{\rm PMNS}$.

The active sterile mixing angle between the light flavour $\alpha$ and the heavy generation $i$ is, according to Eq.~\eqref{eq:p_2}, given by~\cite{Barry:2011fp,Asaka:2011pb}
\begin{equation}
 \theta_{\alpha i} \equiv \overline{U}_{\alpha,3+i} = \left[ m_D^* {M_R^{-1}}^* V_R \right]_{\alpha i},
 \label{eq:p_4}
\end{equation}
and Eq.~\eqref{eq:p_3} as well as the diagonal $M_R$ lead to
\begin{equation}
 \theta_{\alpha i} = - i \left[ U {\rm diag}(\sqrt{m_1}, \sqrt{m_2}, \sqrt{m_3}) R^\dagger {\rm diag}(1/\sqrt{M_1}, 1/\sqrt{M_2}, 1/\sqrt{M_3}) \right]_{\alpha i}.
 \label{eq:p_5}
\end{equation}
A trivial calculation yields:
\begin{equation}
 \theta_{\alpha i} = - i \sum_{k,l,m} U_{\alpha k} \sqrt{m_k} \delta_{kl} R^\dagger_{lm} \frac{1}{\sqrt{M_m}} \delta_{mi} = - i \sum_{k,l} \sqrt{\frac{m_k}{M_i}} U_{\alpha k} \delta_{kl} R^*_{il} = - i \sum_k \sqrt{\frac{m_k}{M_i}} U_{\alpha k} R^*_{ik}.
 \label{eq:p_6}
\end{equation}
Aiming at using the X-ray bound, the observable combinations of the mixing angles take on the form~\cite{Barry:2011fp}
\begin{equation}
 \theta_i^2 \equiv \sum_\alpha |\theta_{\alpha i}|^2 = \sum_\alpha \sum_{k,l} \frac{\sqrt{m_k m_l}}{M_i} U_{\alpha k} U_{\alpha l}^* R^*_{ik} R_{il} = \sum_{k,l} \frac{\sqrt{m_k m_l}}{M_i} \left( \sum_\alpha U_{l \alpha}^\dagger U_{\alpha k} \right) R_{il} R^\dagger_{ki}.
 \label{eq:p_7}
\end{equation}
Using the fact that the PMNS matrix is unitary,\footnote{Note that, in Eq.~\eqref{eq:p_2}, the PMNS matrix $U$ is exactly unitary since it diagonalizes the upper $3\times 3$ block $m_\nu$ of the black-diagonalized version of the full $6\times 6$ matrix $M_\nu$. Non-unitarity of the PMNS matrix~\cite{Antusch:2006vwa} would apply to the upper left $3\times 3$ of the matrix $\overline{U}$, which is only approximately unitary.} we can conclude that $\sum_\alpha U_{l \alpha}^\dagger U_{\alpha k} = \delta_{lk}$, which yields
\begin{equation}
 \theta_i^2 = \sum_k \frac{m_k}{M_i} R_{ik} R^\dagger_{ki}.
 \label{eq:p_8}
\end{equation}
Note that no sum over $i$ is implied in Eq.~\eqref{eq:p_8}.

Now we can see what happens: if the light neutrinos are quasi-degenerate in mass, $m_k\simeq m_0 \gg \sqrt{\Delta m^2_A}$ for $k=1,2,3$, then the factor $m_k$ can be pulled out of the sum in Eq.~\eqref{eq:p_8}. If in addition the complex orthogonal matrix $R$ is also unitary (i.e., it must actually be real,\footnote{From $R^T R = R^\dagger R = \mathbbm{1}$ it follows immediately that $R^* = R$, which means that $R$ is real.} as is the case in many models in the literature), then $\sum_k R_{ik} R^\dagger_{ki} = \delta_{ii} = 1$ implies an interesting correlation between the $i$-th active-sterile mixing angle, the $i$-th heavy neutrino mass $M_i$, and the light neutrino mass $m_0$:
\begin{equation}
 \theta_i^2 = \frac{m_0}{M_i}.
 \label{eq:p_9}
\end{equation}
In particular for $i = 1$, a hard bound on $\theta_i$ exists, roughly given by $\theta_1^2 \lesssim 1.8\cdot 10^{-5} \left( \frac{1~{\rm keV}}{M_1} \right)^5$~\cite{Boyarsky:2009ix}. Applying this bound to Eq.~\eqref{eq:p_9}, one obtains
\begin{equation}
 \frac{m_0}{1~{\rm eV}} \lesssim 1.8\cdot 10^{-2} \left( \frac{1~{\rm keV}}{M_1} \right)^4,
 \label{eq:p_10}
\end{equation}
which can never be fulfilled: even for extremely small values of $M_1 \simeq 1$~keV, which are already excluded by Lyman-$\alpha$ (Ly-$\alpha$) data~\cite{Boyarsky:2008xj}, one would need $m_0 \lesssim 0.018~{\rm eV} < \sqrt{\Delta m^2_A} \simeq 0.050~{\rm eV}$~\cite{Tortola:2012te} to fulfill Eq.~\eqref{eq:p_10}, while the condition for a quasi-degenerate active neutrino spectrum would require $m_0 \gg \sqrt{\Delta m^2_A}$. Larger values of $M_1$ would make the situation even worse, and hence the two properties, $N_1$ as keV sterile neutrino Dark Matter and quasi-degenerate active neutrinos, cannot be fulfilled simultaneously. $\Box$\\

Note that there is an interesting nuance to Eq.~\eqref{eq:p_9}: for quasi-degenerate light neutrinos, the cosmological sum of neutrino masses is given by
\begin{equation}
 \Sigma = m_1 + m_2 + m_3 \simeq 3 m_0,
 \label{eq:p_11}
\end{equation}
in case that we have three light neutrinos contributing to $\Sigma$. This is particularly true in a seesaw type~I setting where the lightest sterile neutrino has a mass of $\mathcal{O}({\rm keV})$. Then, the corresponding correlation looks like:
\begin{equation}
 \theta_i^2 = \frac{\Sigma}{3 M_i}.
 \label{eq:p_12}
\end{equation}
This version of Eq.~\eqref{eq:p_9} is another testable relation, at least in principle, and this time by data from astrophysical observations only. However, note that systematic errors in $\Sigma$ are known to be able to lead to wrong conclusions on the neutrino mass~\cite{Maneschg:2008sf}, which could also lead to problems when applying Eq.~\eqref{eq:p_12}.

\section{\label{sec:disc}Discussion}

\begin{figure}[t]
\centering
\includegraphics[width=12cm]{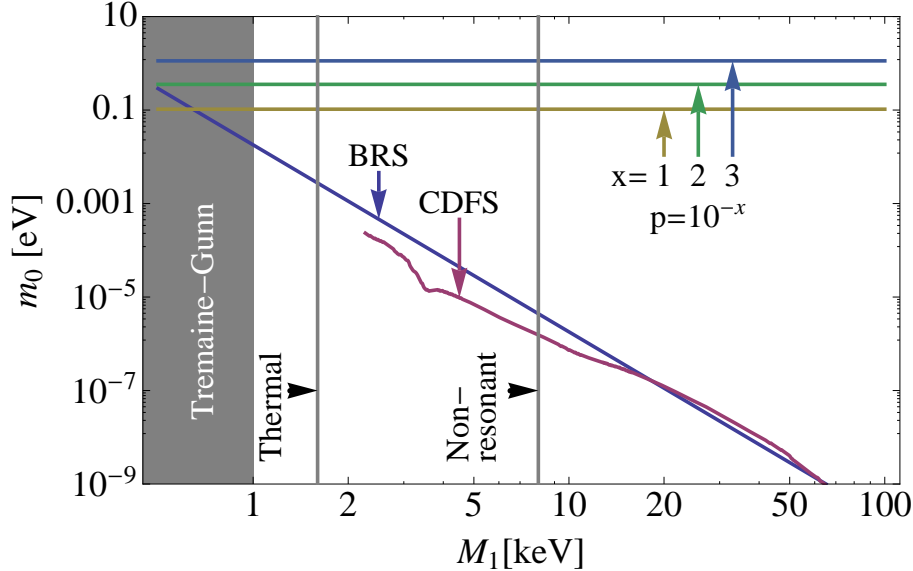}
\caption{\label{fig:NoGo} The illustration of the No-Go theorem in the $m_0$--$M_1$ plane. The requirement of having QD active neutrinos yields a \emph{lower} bound on $m_0$, while the X-ray bound yields an \emph{upper} bound. Clearly, both requirements are inconsistent with each other.}
\end{figure}

Let us graphically illustrate the situation described in Sec.~\ref{sec:theorem}, which is done in Fig.~\ref{fig:NoGo}. We have plotted the \emph{upper} bound on $m_0$ arising from Eq.~\eqref{eq:p_9} when the X-ray bound is imposed. We have used both, the simplified form of the X-ray bound by Boyarsky, Ruchayskiy, and Shaposhnikov (BRS)~\cite{Boyarsky:2009ix}, as done in Eq.~\eqref{eq:p_10}, and the more elaborate bound by Canetti, Drewes, Frossard, and Shaposhnikov (CDFS)~\cite{Canetti:2012vf,Canetti:2012kh}, who combined data from Refs.~\cite{Dolgov:2000ew,Abazajian:2001vt,Boyarsky:2005us,Boyarsky:2006fg,RiemerSorensen:2006fh,Abazajian:2006yn,Watson:2006qb,Boyarsky:2006ag,Abazajian:2006jc,Boyarsky:2007ay,Boyarsky:2007ge,Loewenstein:2008yi}. This has to be compared with the requirement of having a quasi-degenerate mass pattern for active neutrinos, which can be described by the use of a \emph{degeneracy parameter} $p$ defined by
\begin{equation}
 p \equiv \frac{\sqrt{m^2 + \Delta m^2_A}-m}{\sqrt{m^2 + \Delta m^2_A}} \in [0,1],
 \label{eq:d_1}
\end{equation}
where $m$ denotes the lightest active neutrino mass. We have displayed the resulting \emph{lower} bounds for relative degeneracies of $10\%$, $1\%$, and $0.1\%$. For illustration, we have also included some lower bounds on $M_1$, which arise from cosmological structure formation: the Tremaine-Gunn bound~\cite{Tremaine:1979we} applied to the case of keV sterile neutrinos roughly leads to $M_1 \gtrsim 1$~keV~\cite{Boyarsky:2008ju,Gorbunov:2008ka}. More elaborate bounds take into account the details of the production mechanism: with the Lyman-$\alpha$ data as example, for non-resonantly produced keV steriles~\cite{Dodelson:1993je} the lower bound is about $M_1 \gtrsim 8$~keV~\cite{Boyarsky:2008xj}, while for the case of thermal overproduction with subsequent entropy dilution this bound has to be rescaled to $M_1 \gtrsim 1.6$~keV~\cite{Bezrukov:2009th}. Note that generic upper bounds on $M_1$ are in the region of about 50~keV, but the exact value depends on the production mechanism~\cite{Canetti:2012kh,Abazajian:2012ys}.

Let us now discuss how certain models could or could not be probed by the No-Go theorem: first of all, as explained in Sec.~\ref{sec:theorem}, we can only constrain QD situations. However, in mass models for keV sterile neutrinos we need a strong hierarchy in the sterile sector~\cite{Abazajian:2012ys}. In some models, such as the ones based on $L_e - L_\mu - L_\tau$ symmetry~\cite{Shaposhnikov:2006nn,Lindner:2010wr} or the ones based on the Froggatt-Nielsen mechanism~\cite{Barry:2011wb,Merle:2011yv,Barry:2011fp}, these hierarchies tend to translate into the light neutrino sector. In the extended seesaw mechanism~\cite{Barry:2011wb,Zhang:2011vh}, one light neutrino is even exactly massless. On the other hand, in models such as the ones based on the split seesaw mechanism~\cite{Kusenko:2010ik,Adulpravitchai:2011rq} there is no such natural tendency. These latter types of models could be constrained by the No-Go theorem.

A realistic situation in a few years from now is that an experiment like KATRIN~\cite{Osipowicz:2001sq} or MARE~\cite{Nucciotti:2010tx}, ideally in accordance with an experiment like GERDA~\cite{GERDA} proving the Majorana nature of neutrinos, has established the light neutrino mass pattern to be QD, while direct DM searches continued to fail detecting anything. Then, we would need to find models explaining this situation, while at the same time trying to keep a non-WIMP DM candidate like the keV sterile neutrino in existence. In such a situation, as soon as the CI matrix is real in a certain model, the No-Go theorem applies. While the requirement of a real matrix $R$ sounds very restrictive at first sight, it is actually quite often the case that models invoke situations like the absence of $CP$ violation. In such a case, the theorem would immediately be applicable. Indeed, a whole class of $CP$ symmetric models for keV sterile neutrinos would be in trouble in that situation. But even if $CP$ violation is present, it could well be completely contained in the light sector Dirac and Majorana $CP$ phases $\delta$ and $(\alpha, \beta)$ of the PMNS matrix, respectively, while the CI matrix $R$ is still real.

Let us end by discussing how to avoid the situation of the No-Go theorem. The theorem presented here is, as any No-Go theorem, based on certain assumptions. If one of these assumptions is not fulfilled, the theorem is not applicable. In order to illustrate the points more clearly, we make a short list of the assumptions involved and describe how to avoid them:
\begin{itemize}

\item \emph{QD light neutrinos}:

If the light neutrino mass pattern is not QD, the step from Eq.~\eqref{eq:p_8} to Eq.~\eqref{eq:p_9} is invalid. However, at the moment we do not know this mass pattern and we have to wait for experimental input.

\item $R$ \emph{must be real}:

If there is considerable $CP$ violation in the model, ${\rm Im}(R)$ is not negligible. This would also invalidate the step from Eq.~\eqref{eq:p_8} to Eq.~\eqref{eq:p_9}. On the other hand, if a QD pattern is observed and a keV sterile neutrino is the DM in a certain model (or maybe even identified in a lab-based experiment~\cite{Bezrukov:2006cy,He:2009mv,Liao:2010yx,deVega:2011xh}), the combination of all information would proof the existence of $CP$ violation by principle reasons.

\item \emph{The framework must be close to seesaw type~I}:

We need a seesaw type~I situation in order for the form of the matrix in Eq.~\eqref{eq:p_1} to be valid. While any seesaw type~II~\cite{Magg:1980ut,Lazarides:1980nt} contribution destroys that argument, this is not true for a split seesaw setup~\cite{Kusenko:2010ik,Adulpravitchai:2011rq}. Even in more general seesaw scenarios, Eq.~\eqref{eq:p_1} and the subsequent arguments could be valid up to the dimensions of the matrices.

\item $N_1$ \emph{must decay like} $N_1 \to \nu \gamma$:

If for some reason this decay is hindered, e.g.\ by a symmetry that stabilizes $N_1$, then the X-ray bound displayed in Fig.~\ref{fig:NoGo} does not apply. In such a case, there would be no inconsistency in the observables, and the No-Go theorem would not be valid.

\item $N_1$ \emph{must make up a considerable part of the DM}:

If we have multi-species DM, the X-ray bound would not be fully invalidated, but certainly be weakened. This is not a priori a problem, since the plot presented in Fig.~\ref{fig:NoGo} is logarithmic and even a weaker X-ray bound could lead to an inconsistent situation in which the No-Go theorem applies. However, there might be cases in which a sufficiently stable keV sterile neutrino $N_1$ exists but it makes up only a tiny fraction of the DM in the Universe. While in such a situation we might not even talk about ``keV sterile neutrino DM'', it would nevertheless formally invalidate the No-Go theorem.

\end{itemize}
Hence, as for any No-Go theorem, there are ways to avoid it. On the other hand, in situations where it is applicable the theorem could be a powerful handle to constrain or even exclude certain models, by this contributing to our knowledge on neutrinos and Dark Matter.

\section{\label{sec:conc}Conclusions}

We have presented a No-Go theorem which shows that, under certain circumstances, a quasi-degenerate mass pattern of light neutrinos is not consistent with keV sterile neutrino Dark Matter. While there are clearly situations in which the theorem is invalid, it nevertheless applies to the relatively large class of models with a real CI-matrix $R$ in a seesaw type~I situation and an unstable keV sterile neutrino as Dark Matter. After stating and proving the theorem, we have given a discussion of its implications and of the possible ways around it. The theorem could have very interesting consequences for model building in case that experiments on the light neutrino mass reveal at some point a QD mass pattern.

\section*{\label{sec:ack}Acknowledgements}

AM would like to thank S.~F.~King and V.~Niro for carefully reading the manuscript and giving valuable comments. AM acknowledges financial support by a Marie Curie Intra-European Fellowship within the 7th European Community Framework Programme FP7-PEOPLE-2011-IEF, contract PIEF-GA-2011-297557, and partial support from the European Union FP7  ITN-INVISIBLES (Marie Curie Actions, PITN-GA-2011- 289442).

\bibliographystyle{./apsrev}
\bibliography{./keV_NoGo}

\end{document}